\documentclass[12pt, onecolumn, draftcls]{IEEEtran}

\usepackage{graphicx}
\usepackage{url}
\usepackage{epstopdf}
\usepackage{amsfonts,amssymb, amsbsy, amsthm}
\usepackage{amsmath}
\usepackage{times}
\usepackage{tikz}
\usetikzlibrary{arrows, positioning}

\tikzset{
    >=stealth',
    user/.style={
           rectangle,
           rounded corners,
           draw,
           minimum height=2em,
           minimum width=1cm,
           text centered},
    relay/.style={
           rectangle,
           rounded corners,
           draw,
           minimum height=1em,
           minimum width=1cm,
           text centered},
    pil/.style={
           ->,
           shorten <=2pt,
           shorten >=2pt},
    pil_rev/.style={
           <-,
           shorten <=2pt,
           shorten >=2pt},
    pil_dash/.style={
    		->, dashed,
    		shorten <=2pt,
    		shorten >=2pt}
}

\DeclareMathOperator{\Mat}{Mat}
\DeclareMathOperator{\snr}{SNR}
\DeclareMathOperator{\Aut}{Aut}
\DeclareMathOperator{\rk}{rank}
\DeclareMathOperator{\id}{id}
\DeclareMathOperator{\diag}{diag}
\DeclareMathOperator{\Nm}{Nm}

\usepackage[latin1]{inputenc}      

\begin{document}

\title{Constructions of Fast-Decodable Distributed Space-Time Codes}
\author{Amaro Barreal$^\dagger$, Camilla Hollanti$^\dagger$, and Nadya Markin$^\ddag$ \\

\begin{small}
$\dagger$ Department of Mathematics and Systems Analysis, Aalto University, P.O. Box 11100, FI-00076 Aalto, Finland, \texttt{(amaro.barreal, camilla.hollanti)@aalto.fi}
\\
$\ddag$ School of Physical and Mathematical Sciences, Nanyang Technological University, 21 Nanyang Link, Singapore 637371, \texttt{nadyaomarkin@gmail.com}
\end{small}
}
\maketitle

\begin{abstract}
Fast-decodable distributed space-time codes are constructed by adapting the iterative code construction introduced in \cite{Markin} to the $N$-relay multiple-input multiple-output channel, leading to the first fast-decodable distributed space-time codes for more than one antenna per user. Explicit constructions are provided alongside with a performance comparison to non-iterated (non-) fast-decodable codes.
\end{abstract}

\begin{IEEEkeywords} Space-Time Codes, Fast-Decodability, Half-Duplex Relay Channel, Cyclic Division Algebras
\end{IEEEkeywords}

\section{Introduction}
\label{sec:intro}
The increasing interest in \emph{cooperative diversity} techniques as well as rapid growth in the field of multi-antenna communications motivates the investigation of flexible coding techniques for the multiple-input multiple-output (MIMO) cooperative channel. The tools developed in \cite{HollantiRelay1} and \cite{HollantiRelay2} provide the necessary tools to construct fast-decodable space-time (ST) codes for the $N$-relay non-orthogonal amplify-and-forward (NAF) cooperative channel with a single antenna at both the source and the relays. Our work extends these methods to the $N$-relay MIMO NAF channel, that is the relays are allowed to employ multiple antennas for transmission and reception. In addition, many ST codes for the relay scenario exhibit a high rate and hence require a high number of receive antennas at the destination, whereas in this work a single antenna suffices.

\section{The MIMO Amplify-and-Forward Channel}
\label{sec:channel}

We consider the case of a single user communicating with a single destination over a wireless network. $N$ intermediate relays participate in the transmission process. 
In addition, we assume the \emph{half-duplex} constraint, that is the relays can only either receive or transmit a signal in a given time instance. 

Denote by $n_s$, $n_d$ and $n_r$ the number of antennas at the source, destination, and relays, respectively. A superframe consisting of $N$ consecutive cooperation frames of length $T$, each composed of two partitions of $T/2$ symbols, is defined, and all channels are assumed to be static during the transmission of the entire superframe. 

\begin{center}
\fbox{
\begin{minipage}{0.4\textwidth}
\begin{tikzpicture}[node distance=1cm]
 \node[relay] (r1) {\scriptsize Relay 1};
 \node[below=0.15cm of r1] (vert) {\scriptsize $\textbf \vdots$};
 \node[relay, below=0.15cm of vert] (rN) {\scriptsize Relay N};
 \node[above=0.5cm of r1] (dummy) {};
 \node[user, right=1.75cm of dummy] (dest) {\scriptsize Dest.}
   edge[pil_rev] node[pos=0.75, above]{\tiny $G_1$} (r1.east) 
   edge[pil_rev] node[pos=0.65, left]{\tiny $G_N$} (rN.east); 
 \node[user, left=1.75cm of dummy] (source) {\scriptsize Source}
   edge[pil] node[above]{\scriptsize $F$} (dest.west)       
   edge[pil_dash] node[pos=0.75, above]{\tiny $H_1$} (r1.west)
   edge[pil_dash] node[pos=0.65, right]{\tiny $H_N$} (rN.west);
\end{tikzpicture}
\end{minipage}
\hspace{0.3cm}
\begin{minipage}{0.35\textwidth}
	$F$, $H_i$ and $G_i$, $1 \le i \le N$ denote the Rayleigh distributed channels from the source to the destination, relays, and from the relays to the destination, respectively.  
\end{minipage}
}
\end{center}

\vspace{1em}

In a realistic scenario, $n_r \le n_s$, the transmission process can be modeled as
\begingroup
\fontsize{10pt}{12pt}\selectfont
\begin{align*}
	Y_{i,1} &= \gamma_{i,1} F X_{i,1} + V_{i,1}\,, &i=1,\ldots,N \\
	Y_{i,2} &= \gamma_{i,2} F X_{i,2} + V_{i,2} + \gamma_{R_i} G_i B_i(\gamma_{R_i}' H_i X_{i,1} + W_i)\,, &i=1,\ldots,N
\end{align*}
\endgroup
where $Y_{i,j}$ and $X_{i,j}$ are the received and sent matrices, $V_{i,j}$ and $W_i$ represent additive white Gaussian noise, the matrices $B_i$ are needed for normalization and $\gamma_{i,j}$, $\gamma_{R_i}$, $\gamma_{R_i}'$ are signal-to-noise ($\snr$) related scalars. 
 
From the destinations point of view, the above transmission model can be viewed as a virtual single-user MIMO channel as 
\[ Y_{n_d\times n} = H_{n_d\times n} X_{n\times n} + V_{n_d\times n},\]
where $n=N(n_s+n_r)$, $X$ and $Y$ are the (overall) transmitted and received codewords whose structure will take a particular form, and the channel matrix $H$ is determined by the different relay paths. For more details, as well as for the remaining case $n_r > n_s$, we refer to \cite{Yang}.

\subsection{Optimal Space-Time Codes for MIMO NAF Relay Channels} 
\label{subsec:optimal_codes}

Let $\overset{\cdot}{=}$ denote exponential equality, \emph{i.e.}, we write 
\[f(\snr) \overset{\cdot}{=} \snr^b \Leftrightarrow \lim\limits_{\snr\to\infty}{\frac{\log f(\snr)}{\log\snr} = b}\]
and similar for $\overset{\cdot}{\le}$. Consider an $n_d\times n$ MIMO channel and let $\mathcal{A}$ be a scalably dense alphabet \cite[p.651, Definition 2]{Yang}, that is for $\mathcal{A}(\snr)$ its value at a given $\snr$ and for $0 \le r \le \min\lbrace n_d, n\rbrace$ we require
\begin{align*}
	|\mathcal{A}(\snr)| &\overset{\cdot}{=} \snr^{\frac{r}{n}} \\
	|a|^2 &\overset{\cdot}{\le} \snr^{\frac{r}{n}} \text{ for } a \in \mathcal{A}(\snr),
\end{align*}  
for instance PAM or (rotated) QAM constellations. An $n\times n$ ST code $\mathcal{X}$ such that 
	\begin{enumerate}
		\item The entries of any $X \in \mathcal{X}$ are linear combinations of elements in $\mathcal{A}$. 
		
		\item On average, $R$ complex symbols from $\mathcal{A}$ are transmitted per channel use. 
		
		\item $\min\limits_{X_i \neq X_j \in \mathcal{X}}|\det(X_i-X_j)| \ge \kappa > 0$ for a constant $\kappa$ independent of the $\snr$. 
	\end{enumerate}
is called a \emph{rate-$R$ non-vanishing determinant} (NVD) code, and we say the code is \emph{full-rate} if $R = n_d$. This is the largest rate that still allows for the use of a linear decoder, \emph{e.g.}, a sphere decoder, with $n_d$ antennas at the destination.  

Consider an $N$-relay MIMO NAF channel. It was shown in \cite{Yang} that  given a rate-$2n_s$ NVD block-diagonal code $\mathcal{X}$, thus where each $X \in \mathcal{X}$ takes the form $X = \diag\lbrace\Xi_i\rbrace_{i=1}^{N}$ with $\Xi_i \in \Mat(2n_s, \mathbb{C})$, the equivalent code 
\[C = \begin{bmatrix} C_1 & \cdots & C_N\end{bmatrix}, \quad C_i = \begin{bmatrix} \Xi_i\left[1:n_s, 1:2n_s\right] & \Xi_i\left[n_s+1:2n_s,1:2n_s\right]\end{bmatrix}
\] 
achieves the optimal diversity- multiplexing tradeoff (DMT) for the channel, transmitting $C_i$ in the $i^{\text{th}}$ cooperation frame.  

It would thus be desirable to have block-diagonal ST codes which in addition achieve:
\begin{enumerate}
	\item \emph{Full rate $n_d$}: The number of independent complex symbols (\emph{e.g.}, QAM) per codeword equals $n_d(n_s+n_r)N$.  
	
	\item \emph{Full rank}: $\min\limits_{X_i \neq X_j \in \mathcal{X}} \rk(X_i-X_j) = (n_s+n_r)N$. 
	
	\item \emph{NVD}: $\min\limits_{X_i \neq X_j \in \mathcal{X}}|\det(X_i-X_j)|^2 \ge \kappa > 0$ for a constant $\kappa$. 
\end{enumerate}
 The last condition can be abandoned at the low $\snr$ regime without compromising the performance. For very low $\snr$, even relaxing on the full-rank condition does not have adverse effect, since the determinant criterion is asymptotic in nature.

\subsection{On Fast-Decodability}
\label{subsec:decodability}

Consider a ST lattice code $\mathcal{X} = \lbrace \sum_{i=1}^{k}{z_i \cdot B_i} \mid z_i \in \mathbb{Z} \cap J \rbrace$, where $\lbrace B_i \rbrace_{i=1}^{k}$, $k \le 2n^2$, are lattice basis matrices of a rank-$k$ lattice $\Lambda \subseteq  \Mat(n, \mathbb{C})$,  and $J \subset \mathbb{Z}$ is finite and referred to as the \emph{signaling alphabet}. Maximum-likelihood decoding of ST codes amounts to finding the codeword in $\mathcal{X}$ that achieves
\[Z = \arg\min\lbrace ||Y-HX||^2_F \rbrace_{X \in \mathcal{X}}.\] 

Writing $b_i$ for the vectorization of $H B_i$, that is stacking its columns followed by separating the real and imaginary parts, define $B = (b_1,\ldots,b_k)$ and $z = (z_1,\ldots,z_k)^{T}$. Each received codeword can thus be represented as $B\cdot z$. Performing $QR$- decomposition on $B$, where $QQ^{\dagger} = I$, $R$ upper triangular, leads to finding
\begin{small}
\[\arg\min\lbrace||Y-HX||^2_F\rbrace_{X \in \mathcal{X}} \leadsto \arg\min\lbrace ||Q^{\dagger}y-Rz||^2_E\rbrace_{z \in J^k} = \arg\min\lbrace||y'-Rz||^2_E\rbrace_{z \in J^k}, \]
\end{small}
where $y' = Q^{\dagger}y$. This search can be simplified by using a sphere decoder, the complexity of which is upper bounded by that of exhaustive search, \emph{i.e.}, by $|J|^k$. The structure of the matrix $R$ can however reduce the complexity of decoding. A ST code whose decoding complexity is $|J|^{k'}$, $k' < k-1$, due to the structure of $R$ is called \emph{fast-decodable} \cite{Silver}. 
A more extensive review on fast-decodability can be found in \cite{HollantiRelay2}.

\section{Iterated Space-Time Codes}
\label{sec:codes}

Recently, an iterative ST code construction has been proposed in \cite{Markin}. By choosing the maps and elements involved in the construction carefully, the resulting code can inherit some good properties from the original code, such as fast-decodability or full-diversity. This makes the proposed method an interesting tool for constructing bigger codes from well-performing ones. 

Consider a tower of field extensions $\mathbb{Q} \subseteq \mathrm{F} \subseteq \mathrm{K}$, with $\mathrm{F}/\mathbb{Q}$ finite Galois and $\mathrm{K}/\mathrm{F}$ cyclic Galois of degree $n$ with Galois group $\Gamma(\mathrm{K}/\mathrm{F}) = \langle \sigma \rangle$. Let $\gamma \in \mathrm{F}^\times$ be such that $\gamma^i \notin \Nm_{\mathrm{K}/\mathrm{F}}(\mathrm{K}^\times)$, $i = 1,\ldots,n$, and let $\mathcal{C} = (\mathrm{K}/\mathrm{F},\sigma,\gamma) = \bigoplus_{i=0}^{n-1}{e^i \mathrm{K}}$, where $e^n = \gamma$ and $ke = e\sigma(k)$ for all $k \in \mathrm{K}$, be a cyclic division algebra of dimension $n^2$ over its center $\mathrm{F}$. Given $c = \sum_{i=0}^{n-1}{e^i c_i} \in \mathcal{C}$, the representation over its maximal subfield is
\[\lambda: c \mapsto \left[\begin{smallmatrix} c_0 & \gamma\sigma(c_{n-1}) & \cdots & \gamma\sigma^{n-1}(c_1) \\
c_1 & \sigma(c_0) & \cdots & \gamma\sigma^{n-1}(c_2) \\
\vdots & \vdots & \ddots & \vdots \\
c_{n-1} & \sigma(c_{n-2}) & \cdots & \sigma^{n-1}(c_0) 
\end{smallmatrix}\right]\] 

Let $\tau \in \Aut_{\mathbb{Q}}(\mathrm{K})$ such that $\tau \sigma = \sigma \tau$ and commutes also with complex conjugation, $\tau^2 = 1$ and $\tau(\gamma) = \gamma$. Fix $\theta \in \mathrm{F}^\times$ such that $\tau(\theta) = \theta$. Setting $X = \lambda(x)$, $Y = \lambda(y) \in \Mat(n,\mathcal{C})$, define a map
\[\alpha_{\theta}: (X,Y) \mapsto \begin{bmatrix} X & \theta\tau(Y) \\ Y & \tau(X)\end{bmatrix}.\]

The conditions imposed on $\tau$ and $\theta$ ensure that the image of $\alpha_{\theta}$ is an $\mathrm{F}$-algebra, and is division if and only if $\theta \neq c \tau(c)$ for all $c \in \mathcal{C}$. For more details on this construction method, the reader may consult \cite{Markin}.

\section{Distributed Iterated Space-Time Codes}
\label{sec:stc_relay}

Consider an $N$-relay MIMO channel. Given a ST code $\mathcal{X} \subset \Mat(n,\mathbb{C})$, we define the following map $f: \mathcal{X} \to \Mat(nN, \mathbb{C})$
\[f_{\eta}^N: X \mapsto \diag\lbrace\eta^i(X)\rbrace_{i=0}^{N-1}\]  
for a suitable function $\eta$ such that $\eta^{N} = \id$. In the following, we will make use of this function to construct distributed ST codes from iterated and non-iterated codes. Often the map $\eta$ is chosen to be a field automorphism, so that the determinant will correspond to a field norm and be non-vanishing with a suitable choice of fields. Here, we choose $\eta = \id$, as in our specific examples the blocks composing the codeword matrices will already have the NVD property, thus no special modifications will be necessary. Note that if the blocks $\eta^i(X)$ are fast-decodable, the resulting block-diagonal code will also be fast-decodable \cite{HollantiRelay2}.

\subsection{Explicit Constructions for $N = 2$ Relays and $n_r + n_s = 4$}
\label{subsec:performance}

In the following, let $N = 2$ be the number of relays, both equipped with $n_r$ antennas and such that $n_r+n_s = 4$. We construct three different codes, each of them with different characteristics, arising from the following towers of extensions: 

\begingroup
\fontsize{08pt}{10pt}\selectfont
\begin{minipage}{0.3\textwidth}
\centering
\begin{tikzpicture}[node distance=0.25cm]
 \node (s_alg) {$\mathcal{C}_s = (K_s/F_s,\sigma_s,-1)$};
 \node[below=0.35cm of s_alg] (s_k) {$K_s = F_s(i)$}
 	edge[-] node[pos=0.5, right]{2} (s_alg);
 \node[below=0.35cm of s_k] (s_f) {$F_s = \mathbb{Q}(\sqrt{-7})$}
 	edge[-] node[pos=0.5, right]{2} (s_k);
 \node[below=0.35cm of s_f] (s_q) {$\mathbb{Q}$}
 	edge[-] node[pos=0.5, right]{2} (s_f);
\end{tikzpicture}

\centering
$\sigma_s: i \mapsto -i$ 
\end{minipage}\hfill
\begin{minipage}{0.3\textwidth}
\centering
\begin{tikzpicture}[node distance=0.35cm]
 \node (g_alg) {$\mathcal{C}_g = (K_g/F_g,\sigma_g,i)$};
 \node[below=0.35cm of s_alg] (g_k) {$K_g = F_g(\sqrt{5})$}
 	edge[-] node[pos=0.5, right]{2} (g_alg);
 \node[below=0.35cm of s_k] (g_f) {$F_g = \mathbb{Q}(i)$}
 	edge[-] node[pos=0.5, right]{2} (g_k);
 \node[below=0.35cm of s_f] (g_q) {$\mathbb{Q}$}
 	edge[-] node[pos=0.5, right]{2} (g_f);
\end{tikzpicture}

$\sigma_g: \sqrt{5} \mapsto -\sqrt{5}$
\end{minipage}\hfill
\begin{minipage}{0.3\textwidth}
\centering
\begin{tikzpicture}[node distance=0.35cm]
 \node (m_alg) {$\mathcal{C}_m = (K_m/\mathbb{Q},\sigma_m,-\frac{8}{9})$};
 \node[below=0.35cm of m_alg] (m_k) {$K_m = \mathbb{Q}(\zeta_5)$}
 	edge[-] node[pos=0.5, right]{2} (m_alg);
 \node[below=1.2cm of s_k] (m_q) {$\mathbb{Q}$}
 	edge[-] node[pos=0.5, right]{4} (m_k);
\end{tikzpicture}

$\sigma_m: \zeta_5 \mapsto \zeta_5^3$ 
\end{minipage}
\endgroup

\begin{enumerate}
\item The \emph{Silver code}, well known to be fast-decodable \cite{Silver}, is constructed from the cyclic division algebra $\mathcal{C}_s$ and is a finite subset of
\[\Bigg\lbrace  \frac{1}{\sqrt{7}}\left[\begin{smallmatrix} x_1\sqrt{7}+(1+i)x_3 + (-1+2i)x_4 & - x_2^\ast\sqrt{7}-(1-2i)x_3^\ast-(1+i)x_4^\ast \\ x_2\sqrt{7}-(1+2i)x_3-(1-i)x_4 & x_1^\ast \sqrt{7} - (1-i)x_3^\ast - (-1-2i)x_4^\ast \end{smallmatrix}\right] \Bigg\vert x_j \in \mathbb{Z}[i],1 \le j \le 4\Bigg\rbrace.\]

Choosing $\theta_s = -17$, $\tau_s = \sigma_s$, and given set elements $X = X(x_1,x_2,x_3,x_4)$, $Y = Y(y_1,y_2,y_3,y_4)$, we construct a distributed iterated Silver code via the map
\[f(\alpha_{\theta_s}(X,Y))_{\id}^{2} = \left[\begin{smallmatrix} \alpha_{\theta_s}(X,Y) & 0 \\ 0 & \alpha_{\theta_s}(X,Y) \end{smallmatrix}\right] = \left[\begin{smallmatrix}X & \theta_s\tau_s(Y) & 0&0  \\ Y & \tau_s(X) & 0& 0\\ 0& 0& X & \theta_s \tau_s(Y) \\  0&0 & Y & \tau_s(X) \end{smallmatrix}\right] \in \Mat(8,K_s).\] 

Set $\mathcal{X}_s = \lbrace \sum_{j=1}^{16}{z_j\cdot S_j} \mid z_j \in J\cap\mathbb{Z}\rbrace$, where a lattice basis $\Lambda_s = \lbrace S_j \rbrace_{j=1}^{16}$ is given by 
\begin{small}
\begin{eqnarray}
\lbrace f(\alpha_{\theta_s}(X(1,0,0,0),Y(0,0,0,0)))_{id}^2,\ldots ,f(\alpha_{\theta_s}(X(0,0,0,0),Y(0,0,0,1)))_{id}^2, \nonumber \\ 
f(\alpha_{\theta_s}(X(i,0,0,0),Y(0,0,0,0)))_{id}^2,\ldots ,f(\alpha_{\theta_s}(X(0,0,0,0),Y(0,0,0,i)))_{id}^2\rbrace. \nonumber
\end{eqnarray}
\end{small}

\item The \emph{Golden code}, a well-performing ST code introduced in \cite{Golden}, is constructed from $\mathcal{C}_g$ and consists of codewords taken from the set
\[\Bigg\lbrace \frac{1}{\sqrt{5}}\left[\begin{smallmatrix} \nu(x_1+x_2\omega) & \nu(x_3 + x_4\omega) \\ i\sigma_g(\nu)(x_3 + x_4 \sigma_g(\omega)) & \sigma_g(\nu)(x_1 + x_2 \sigma_g(\omega)) \end{smallmatrix}\right] \Bigg\vert x_j \in \mathbb{Z}[i], 1 \le j \le 4 \Bigg\rbrace,\]
where $\omega = (1+\sqrt{5})/2$ and $\nu = 1+i-i\omega$. The Golden code, although very good in performance, is not fast-decodable without modifying the sphere decoder used and has higher decoding complexity than the Silver code. 

We set $\theta_g = 1-i$, $\tau_g = \sigma_g$. Then, for two elements $X = X(x_1,x_2,x_3,x_4),Y = Y(y_1,y_2,y_3,y_4)$, the distributed iterated Golden code is constructed as
\[f(\alpha_{\theta_g}(X,Y))_{\id}^{2} = \left[\begin{smallmatrix} \alpha_{\theta_g}(X,Y) & 0 \\ 0 & \alpha_{\theta_g}(X,Y) \end{smallmatrix}\right] = \left[\begin{smallmatrix}X & \theta_g\tau_g(Y) &0 & 0 \\ Y & \tau_g(X) & 0&0 \\ 0&0 & X & \theta_g \tau_g(Y) \\ 0 & 0& Y & \tau_g(X) \end{smallmatrix}\right] \in \Mat(8,K_g).\] 

Set $\mathcal{X}_g = \lbrace\sum_{j=1}^{16}{z_j\cdot G_j} \mid z_j \in J \cap \mathbb{Z} \rbrace$, where a lattice basis $\Lambda_g = \lbrace G_j \rbrace_{j=1}^{16}$ is
\begin{small}
\begin{eqnarray}
\lbrace f(\alpha_{\theta_g}(X(1,0,0,0),Y(0,0,0,0)))_{id}^2,\ldots ,f(\alpha_{\theta_g}(X(0,0,0,0),Y(0,0,0,1)))_{id}^2, \nonumber \\ 
f(\alpha_{\theta_g}(X(i,0,0,0),Y(0,0,0,0)))_{id}^2,\ldots ,f(\alpha_{\theta_g}(X(0,0,0,0),Y(0,0,0,i)))_{id}^2\rbrace. \nonumber
\end{eqnarray}
\end{small}

\item Finally we also consider the fast-decodable MIDO$_{A4}$ code constructed in \cite{mido}, using $\mathcal{C}_m$ as the algebraic structure. Write $\zeta = \zeta_5$ and choose $\lbrace 1-\zeta,\zeta-\zeta^2,\zeta^2-\zeta^3,\zeta^3-\zeta^4\rbrace$ a basis of $\mathbb{Z}[\zeta]$. Setting $r = |-8/9|^{1/4}$, codewords are taken from 
\[\left\{ \left[\begin{smallmatrix} x_1 & -r^2 x_2^\ast & -r^3\sigma_m(x_4) & -r\sigma_m(x_3)^\ast \\ r^2x_2 & x_1^\ast & r\sigma_m(x_3) & -r^2\sigma_m(x_4)^\ast \\ rx_3 & -r^3x_4^\ast & \sigma_m(x_1) & -r^2\sigma(x_2)^\ast \\ r^3x_4 & rx_3^\ast & r^2\sigma_m(x_2) & \sigma_m(x_1)^\ast \end{smallmatrix}\right] \Bigg\vert x_j \in \mathbb{Z}[\zeta], 1\le j \le 4 \right\},\] 
where for $1 \le j \le 4$, $x_j = x_j(l_{4j-3},l_{4j-2},l_{4j-1},l_{4j}) = l_{4j-3}(1-\zeta)+l_{4j-2}(\zeta-\zeta^2)+l_{4j-1}(\zeta^2-\zeta^3)+l_{4j}(\zeta^3-\zeta^4)$.  
Given an element $X = X(x_1,x_2,x_3,x_4)$ from this set, the adaptation to the cooperative channel is 
\[f(X)_{\id}^2 = \left[\begin{smallmatrix} X & 0 \\ 0 & X \end{smallmatrix}\right] \in \Mat(8,K_m).\]

Set $\mathcal{X}_m = \lbrace \sum_{j=1}^{16}{z_j\cdot M_j} \mid z_j \in J \cap \mathbb{Z} \rbrace$. A lattice basis $\Lambda_m = \lbrace M_j \rbrace_{j=1}^{16}$ is
\begin{small}\[\lbrace X(x_1(1,0,0,0),0,0,0),\ldots ,X(0,0,0,x_4(0,0,0,1))\rbrace.\]\end{small}
\end{enumerate}

\subsection{Determinant and Performance Comparison}
\label{subsec:results}

For the carried out simulations we fix $J = \lbrace \pm 1 \rbrace$, the $2$-PAM signaling constellation. Further, comparison between the constructed codes requires some kind of normalization, and we choose to normalize the volume of the fundamental parallelotope of the underlying lattices to be $\delta(\Lambda) = 1$. We can then compare the distribution of the normalized determinants among all codewords, as illustrated below. In addition to the previously introduced codes, we further consider a modified version of the distributed iterated Silver code using $\theta_s = -1$. Although this choice does not guarantee full-diversity in general, with $2$-PAM the resulting code is still fully diverse. 
\vspace{0.3cm}

\begin{center}
\begingroup
\begin{footnotesize}
\begin{tabular}{|l||c|c|c|c|}
\hline
 & \textbf{Golden} & \textbf{Silver$_{-17}$} & \textbf{Silver$_{-1}$} & \textbf{MIDO$_{A4}$}  \\
 \hline \hline 
Minimum det. & $\pmb{4.445\cdot 10^{-3}}$ & $1.553\cdot 10^{-5}$ & $4.16\cdot 10^{-4}$ & $3.871\cdot 10^{-7}$ \\ 
\hline
Maximum det. & $13.871$ & $4.099$ & $14.268$ & $\pmb{80.500}$ \\
\hline
Average det. & $1.819$ & $0.493$ & $2.007$ & $\pmb{7.485}$ \\
\hline
\end{tabular}
\end{footnotesize}
\endgroup
\end{center}

\begin{figure}[h]	
\begin{center}
	\includegraphics[width=0.5\textwidth]{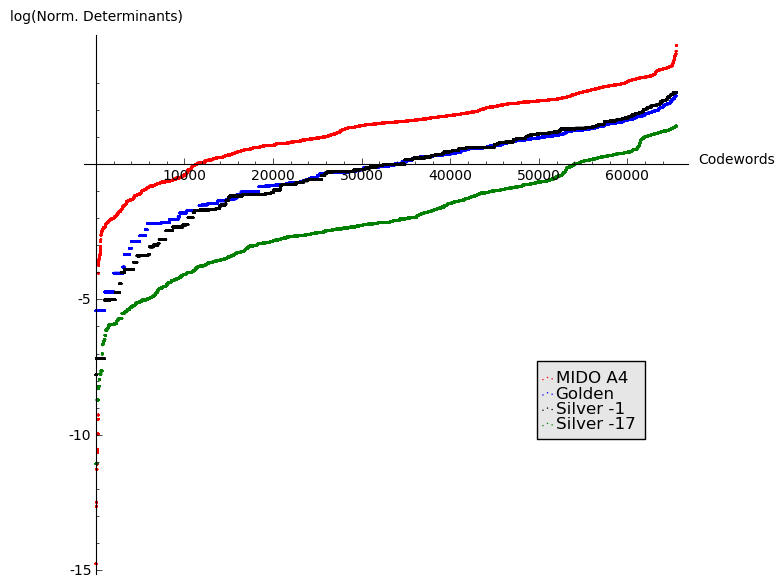}
	\caption{The logarithmic distribution of the normalized determinants of all the $2^{16}$ codewords in $\mathcal{X}_g$, $\mathcal{X}_m$ and $\mathcal{X}_s$ for both $\theta_s = -17$ and $\theta_s = -1$.}
\end{center}
\end{figure}

\begin{figure}[h]
\begin{center}
	\includegraphics[width=0.5\textwidth]{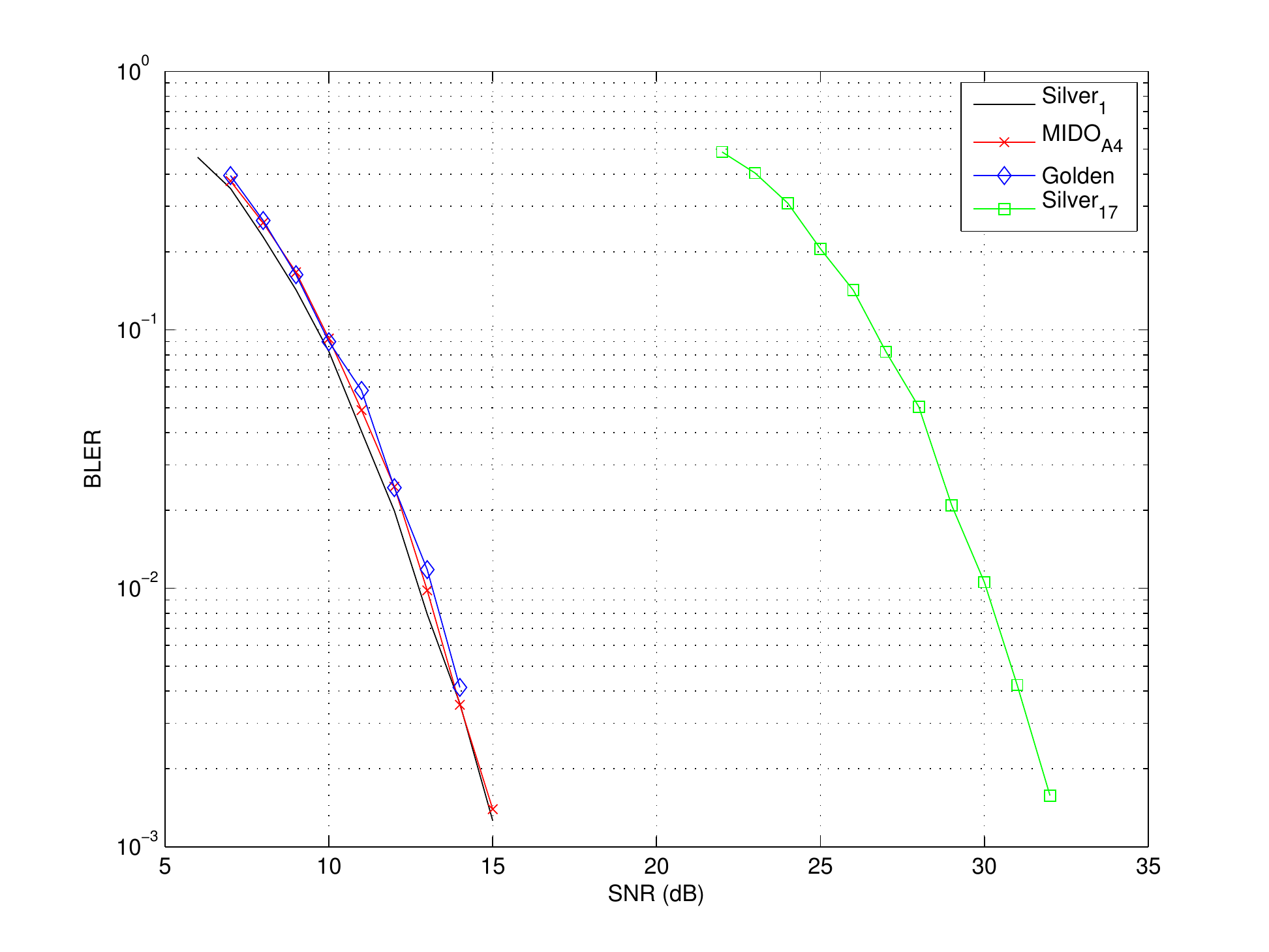}
	\caption{Performance comparison of the above four example codes with $2$-PAM signaling. The data rate is 16/8=2 bits per channel use (bpcu). The Silver code with $\theta_s=-17$ has the worst performance, which is to be expected due to high peak-to-average power ratio stemming from the fact that $|17|$ is not close to one. The other codes perform more or less equally.}
\end{center}
\end{figure}
  
The exact complexity reduction of the iterated distributed codes remains to be examined. It is also not necessarily obvious, that the proposed construction achieves the DMT, since the conditions in \cite{Yang} require that the code rate is $2n_s$, while our example constructions all have code rate $n_d = 1 < 2n_s$. However, since they are full-rate (similarly to the codes in \cite{Yang}) for $n_d$ antennas at the destination, we expect that they do achieve the DMT.

\section*{Acknowledgements}
N. Markin was supported by the Singapore National Research Foundation under Research Grant NRF-RF2009-07

\end{document}